# Wine quality rapid detection using a compact electronic nose system: application focused on spoilage thresholds by acetic acid

**Juan C. Rodriguez Gamboa[a], Eva Susana Albarracin E.[a], Adenilton J. da Silva[b], Luciana Leite[c], Tiago A. E. Ferreira[a]**

[a]*Departamento de Estatística e Informática, Universidade Federal Rural de Pernambuco, Recife, PE, Brazil*

[b]*Centro de Informática, Universidade Federal de Pernambuco, Recife, PE, Brazil*

[c]*Departamento de Tecnologia Rural. Universidade Federal Rural de Pernambuco, Recife, PE, Brazil*

**Abstract**

It is crucial for the wine industry to have methods like electronic nose systems (E-Noses) for real-time monitoring thresholds of acetic acid in wines, preventing its spoilage or determining its quality. In this paper, we prove that the portable and compact self-developed E-Nose, based on thin film semiconductor ($SnO_2$) sensors and trained with an approach that uses deep Multilayer Perceptron (MLP) neural network, can perform early detection of wine spoilage thresholds in routine tasks of wine quality control. To obtain rapid and online detection, we propose a method of rising-window focused on raw data processing to find an early portion of the sensor signals with the best recognition performance. Our approach was compared with the conventional approach employed in E-Noses for gas recognition that involves feature extraction and selection techniques for preprocessing data, succeeded by a Support Vector Machine (SVM) classifier. The results evidence that is possible to classify three wine spoilage levels in 2.7 seconds after the gas injection point, implying in a methodology 63 times faster than the results obtained with the conventional approach in our experimental setup.

*Keywords*: beverage quality control; wine spoilage; online early detection.

1. Introduction

Wine flavor depends on 20 or more compounds, besides water and ethanol, that with subtle alterations in concentration determine its quality (Jackson, 2008). The most important technique used to determine wine quality is directly related to the organoleptic characteristics evaluation by trained experts (Aleixandre,

Cabellos, Arroyo, & Horrillo, 2018; Cretin, Dubourdieu, & Marchal, 2018; Sáenz-Navajas et al., 2015). Since the analytical panels are expensive, time-consuming, and they are not always available, the wine is also characterized using gas and liquid chromatography or spectrophotometry, that require on reagents and experienced personnel (Martins et al., 2018; Perestrelo, Rodriguez, & Câmara, 2017; Stupak, Kocourek, Kolouchova, & Hajslova, 2017; Vazallo-Valleumbrocio, Medel-Marabolí, Peña-Neira, López-Solís, & Obreque-Slier, 2017). Besides, E-Noses are used as an alternative to traditional methods for wines discrimination regarding the organoleptic characteristics. Their purpose is to analyze aroma profiles by registering signals produced by the mixture of gases (as the human nose does) and then comparing the pattern of responses generated by different samples (Lozano, Santos, & Horrillo, 2016; Peris & Escuder-Gilabert, 2016; Rodríguez-Méndez et al., 2016; Zhao et al., 2017). However, most E-Noses are designed for general purpose, and sometimes they are not portable to use on-site.

Volatile acidity (VA) measurements, generally interpreted as acetic acid content ($g \cdot l^{-1}$), are used routinely as an indicator of wine spoilage (Zoecklein, B. W., Fugelsang, K. C., Gump, B. H., & Nury, 1995). Thereby, it is crucial for the wine industry and consumers to have methods for real-time monitoring of VA thresholds. There are previous works which the wine spoilage was characterized using E-Noses developed with special sensors or combined with other technologies and methods. Some common characteristics of those systems are the instrumentation complexity, most of them involve the use of additional equipment that requires experienced personnel, and they do not realize online detection. For instance, a metalloporphyrin based optoelectronic nose was developed in (Amamcharla & Panigrahi, 2010) for the simultaneous prediction of Volatile Organic Compounds (VOCs) concentrations in binary mixtures (acetic acid and ethanol) using partial least square regression (PLSR) and multilayer perceptron neural network (MLP-NN). Besides, in (Gil-Sánchez et al., 2011), it is reported the wine spoilage analysis when in contact with air using a combined system of a potentiometric electronic tongue and a humid E-Nose.

The acetic acid detection was studied by (Macías et al., 2012) using a commercial E-Nose for general purpose, in combination with a neural network classifier (MLP). They detected only the excessive concentrations of acetic acid, equal to or greater than $2g \cdot l^{-1}$ in synthetic wine samples (aqueous ethanol solution at 10% v/v). However, levels higher than $1.2g \cdot l^{-1}$ of VA cause that the wine takes on vinegar aromas (unpleasant), reducing its quality; hence the governments forbid their commercialization (Normative instruction N° 14, 2018; Zoecklein, B. W., Fugelsang, K. C., Gump, B. H., & Nury, 1995). Thus, our work was aimed to detect lower levels and with a quick identification in real wine samples with several spoilage thresholds using the self-developed E-Nose, without using any reagent to reduce the environmental impact, as well with a smooth and safe operation interface (no occupational risk for the operator and with minimal training).

This study presents the self-developed E-Nose based on commercially available gas sensors for early detection of spoilage thresholds by VA in routine tasks of wine quality control. We recorded electrical signals corresponding to odorant profiles of wines samples with different spoilage levels. Afterward, we compared the conventional data processing approach used in E-Noses against our online data processing approach to accelerate the responses. In the conventional approach was applied the preprocessing and feature extraction before an SVM classifier to obtain the main odorant parameters (which requires that the measurement process had finished before data processing stage). By contrast, we focused on an online solution, that let to achieve faster results, using an early portion of the signals while the measurement process is still running. Our approach is based on the training of a deep MLP classifier using the raw data.

2. **Materials and methods**

*2.1 Electronic Nose*

We used an E-Nose, that we named O-NOSE, comprising principally of an array of six metal-oxide gas sensors (**Table 1**), used to detect the volatile compounds. **Fig. 1** shows O-NOSE on the left side, and the sensors board with two layers for a compact design on the right side.

**Table 1.** Gas sensors array setup. The sensors manufactured by Hanwei Sensors[1] are commercially available. They have been chosen because of their high sensitivity to organic, natural, ethanol, methanol, and combustible gases, as well as its simplicity of use and low financial cost.

---

[1] www.hwsensor.com

| Number | Sensor | Description | Load resistance |
|--------|--------|-------------|-----------------|
| 1, 4 | MQ-3 | High sensitivity to alcohol and small sensitivity to Benzine | 22kΩ |
| 2, 5 | MQ-4 | High sensitivity to CH4 and natural gas | 18 kΩ |
| 3, 6 | MQ-6 | High sensitivity to LPG, iso-butane, propane | 22 kΩ |

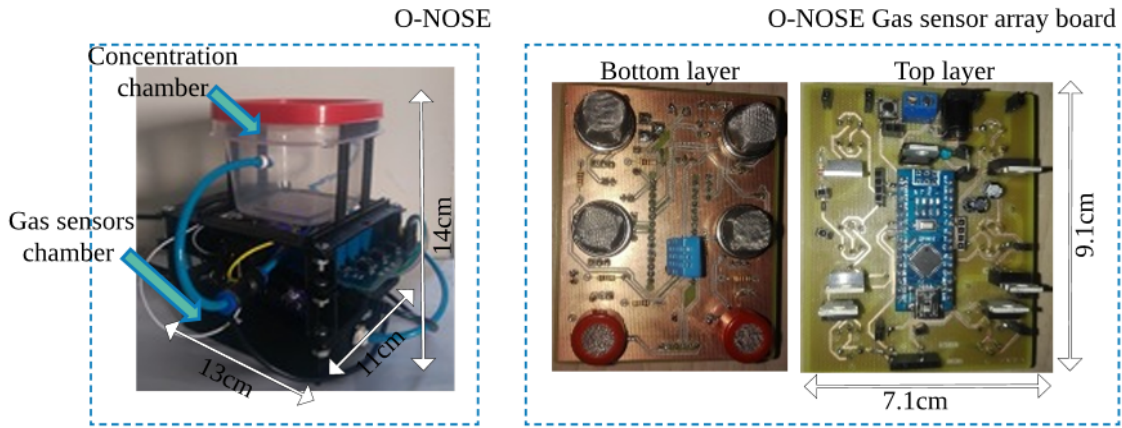

**Fig. 1.** O-NOSE system. On the left side: system appearance and dimensions. Into 100 ml concentration chamber is placed the wine sample. The sensors array is into the 200 ml chamber. On the right side: the main board with the gas sensors and the microcontroller. The gas is sensed by its effect on the sensitive layer of tin dioxide ($SnO_2$), resulting from changes in conductivity brought about by chemical reactions on the surface of the tin dioxide particles.

*2.1.1 Experimental setup*

In **Fig. 2a**, we depict the O-NOSE measurement process divided into three stages. (i) Concentration stage: we used 1ml wine samples to accumulate the volatiles for 30 seconds inside the concentration chamber. (ii) Data acquisition: ten seconds after the initialization of this stage, the VOCs push toward the sensors chamber for 80 seconds generating change in the sensor resistance (gas absorption). Subsequently, the gas injection stops, and it begins the desorption for 90 seconds. Therefore, the acquired data corresponds to 180 seconds with 18.5Hz sample rate. (iii) Purge: the goal is to clean and remove volatile residues for 600 seconds. **Fig. 2b** shows the standard block diagram for the whole experiments, the electrical signals acquired are processed using the pattern recognition techniques after finished the data acquisition stage in the conventional approach or online applying our approach.

*2.2 Data*

*2.2.1 Wine samples*

We used 22 bottles of commercial wines, and to obtain spoiled samples, 13 of the 22 bottles were randomly selected, opened and left in an uncontrolled environment six months before starting the measurements. These bottles were labeled as low-quality (LQ) wines. Besides, another four bottles were opened two weeks before beginning the data collection. These four bottles were labeled as average-quality (AQ) wines, and the remaining five bottles were labeled as high-quality (HQ) wines.

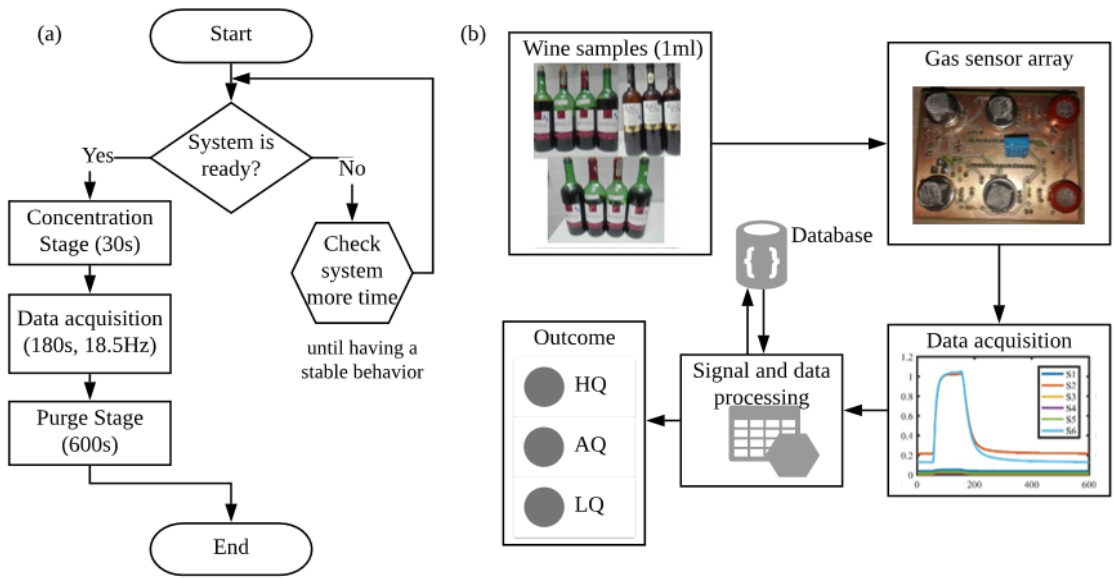

**Fig. 2.** (a) Flowchart of the measurement setup. (b) Block diagram for wine spoilage detection using 1ml samples, the outcome is according to the wine quality.

The 22 wine bottles were characterized as follows: (i) the VA quantification was performed in triplicate according to official methods for wine analysis of the International Organization of Vine and Wine (OIV. International Organization of vine and wine, 2014). (ii) Acetic acid was identified by High Performance Liquid Chromatography (HPLC) with UV/Vis absorption detector, following the procedure detailed in (De Andrade Lima et al., 2010), and the ranges obtained are shown in **Table 2**. It is known that at normal levels in wines (<0.3g·l$^{-1}$) the VA can be a desirable flavor, adding to the complexity of taste and odor, as well, a content of less than 0.70 g·l$^{-1}$ seldom imparts spoilage character. However, a progressive increment in VA gives to the wines a sour taste and taints its fragrance (Jackson, 2008; Zoecklein, B. W., Fugelsang, K. C.,

Gump, B. H., & Nury, 1995). Brazilian Ministry of Agriculture, Livestock and Supply (Instrução Normativa N° 14, 2018) establishes that the maximum level of VA in wine is 1.2 g·l$^{-1}$.

**Table 2.** Ranges detected of volatile acidity and acetic acid according to the wine spoilage thresholds. The ranges presented correspond to the minimum and maximum values of the analysis.

| Wine quality level | Volatile acidity in g·l$^{-1}$ | Acetic acid in g·l$^{-1}$ |
| --- | --- | --- |
| HQ | [0.15, 0.3] | [ND, 0.23] |
| AQ | [0.31, 0.41] | [0.24, 0.34] |
| LQ | [0.8, 3] | [0.74, 2.75] |

ND: not detected.

The database collected using O-NOSE has 235 wines measurements as follow: 51, 43, and 141 measurements of HQ, AQ, and LQ respectively. Besides, we collected 65 ethanol measurements in concentrations (v/v) of 2, 5, 10, 20, 30, and 40 ml of ethanol diluted in distilled water to make solutions of 200ml.

*2.3 Feature extraction and selection*

The most common groups of characteristics extracted from the gas sensors signals are the steady and transient state features (J. Yan et al., 2015). We used 23 features to capture the dynamic and static behavior of each gas sensor. So, we obtained a 138 columns characteristics matrix, where each row represents the fingerprint of one measurement. One example of the raw data (**Fig. 3a**) evidences the sensor sensitivity regarding VOCs analyzed. In **Fig. 3b-c,** we show the steady and transient features for the response of one sensor during the three intervals of the acquisition procedure explained at the end of Section 2.1.

Afterward, we applied the SVM Recursive Feature Elimination Cross Validation (RFECV) method to reduce the dimensionality, looking to generate parsimonious and robustness models (Lin et al., 2012; K. Yan & Zhang, 2015). Thus, it was chosen the followings steady-state characteristics: $\Delta G = max_k g[k] - min_k g[k]$, defined as the maximal conductance change concerning the baseline, and its normalized version ($\|\Delta G\| = (max_k g[k] - min_k g[k])/min_k g[k]$), as well, the area under the curve in the absorption and desorption portions of the gas, blue and gray areas in **Fig. 3b**, respectively. Additionally, we had an aggregate of features reflecting the dynamics of the rising/falling transient portion of the sensor response using an exponential moving average filter (ema$_\alpha$) that converts the transient portion into a real scalar by estimating the maximum/minimum value $y[k] = (1-\alpha)y[k-1] + \alpha(x[k] - x[k-1])$, where $[k = 1, 2, \ldots, T]$, $y[0]$ its initial condition, set to zero ($y[0] = 0$, and the scalar $\alpha(\alpha \in \{0,1\})$ being a smoothing parameter of the operator such as was defined in (Muezzinoglu et al., 2009; Vergara et al., 2012). We tested three different values for $\alpha = 0.1$, $\alpha = 0.01$, and $\alpha = 0.001$ as shown in **Fig. 3c**; and by RFECV feature selection, it was chosen the **max** ema$_\alpha$ with $\alpha = 0.01$ as an informative transient feature.

*2.4 Classification methods*

We use two approaches for the classification tasks in this application. The first one consists in applying feature extraction and selection before the classifier. And the second one consists in processing an early portion of the raw data.

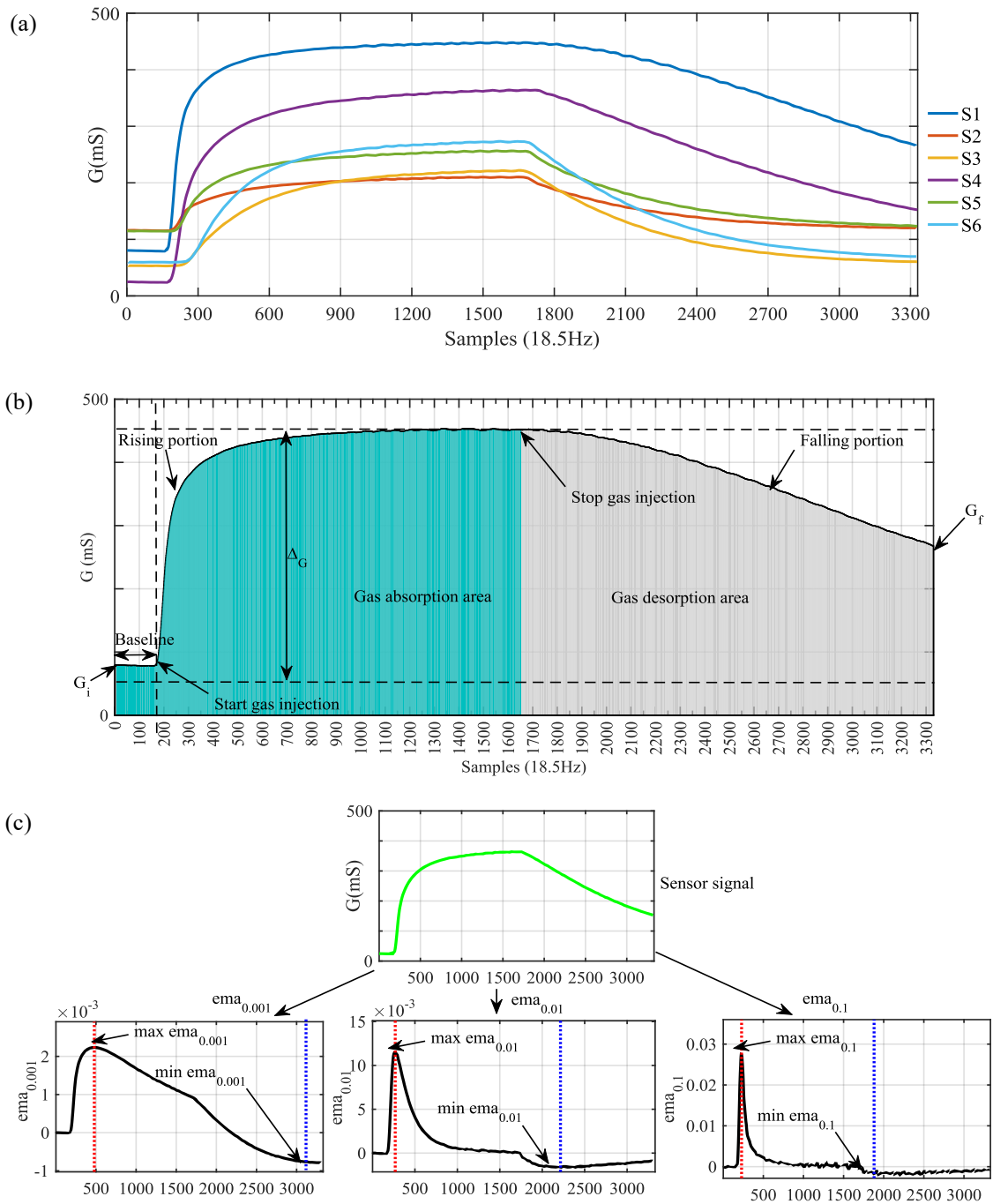

**Fig. 3.** (a) Wine measurement acquired with O-NOSE; S1, S2,…, S6: gas sensor outputs in conductance units G. (b) Output of a gas sensor; Gi: initial conductance value, Gf : final conductance value, $\Delta G$: maximal conductance change concerning the baseline. (c) Dynamics of the rising/falling transient portion using an exponential moving average filter (ema$\alpha$) for $\alpha$=0.1, $\alpha$=0.01, and $\alpha$=0.001.

*2.4.1 Conventional approach to classification using SVM*

In this approach, it is necessary to have the whole measurement to obtain the main odorant parameters. We tested various kernels on an SVM classifier and selected a gaussian kernel; then it was trained the model. The block diagram of this approach (depicted in **Fig. 4**) exhibits the steps performed that includes a feature extraction block generating the $C_{i,j}$ vector, where $i = 1,2 \ldots, 23$ is the number of characteristics and $j = 1,2 \ldots, 6$ is the number of sensors. Afterward, the characteristics vector feed the feature selection block, and finally, the chosen variables are carried to the inputs of the SVM classifier.

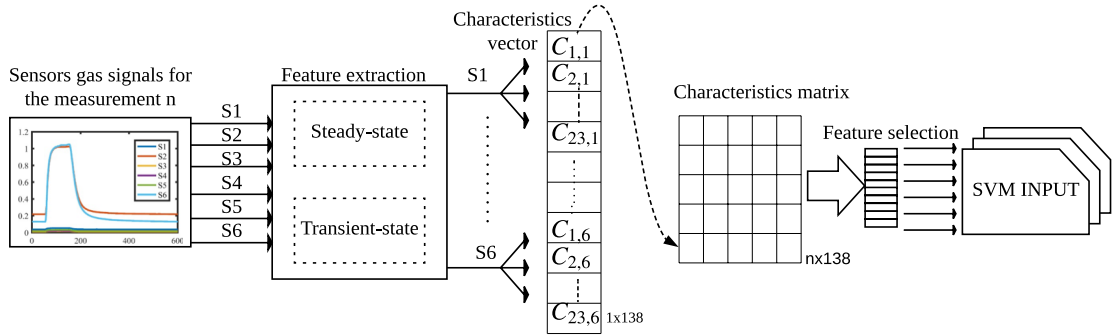

**Fig. 4.** Block diagram of the conventional approach to classification using SVM. This diagram comprises a Feature Extraction block (FE), a Feature Selection block (FS), and subsequently, the characteristics matrix feeds an SVM classifier.

*2.4.2 Rapid and online detection approach using deep MLP*

This approach is based on a neural network classifier that is feed with the raw data to perform the discrimination tasks (Peng, Zhao, Pan, & Ye, 2018). Inspired by the mentioned approach and looking to accelerate the response, we propose a rapid detection method in wine quality control, focused on an online solution that lets to achieve faster results using only an early portion of the signals, similar to the presented in (Längkvist, Coradeschi, Loutfi, & Balaguru Rayappan, 2013) for a meat spoilage application, but using a supervised method: deep MLP neural network. The goal with this approach is to offer the possibility to make estimations a few seconds after beginning the measurement process while it is still running. Note that we did not consider the baseline of the sensor since generally in this slice there is no change. Consequently, the data processing starts instantly before the gas injection. A rising window method was applied to find the minor portion of information with the best performance of the classifier. This reduces the effort to obtain discrimination models since as complicated preprocessing techniques no need to be applied and it is feasible by the computational acceleration in the last years. **Fig. 5** depicts the approach employed.

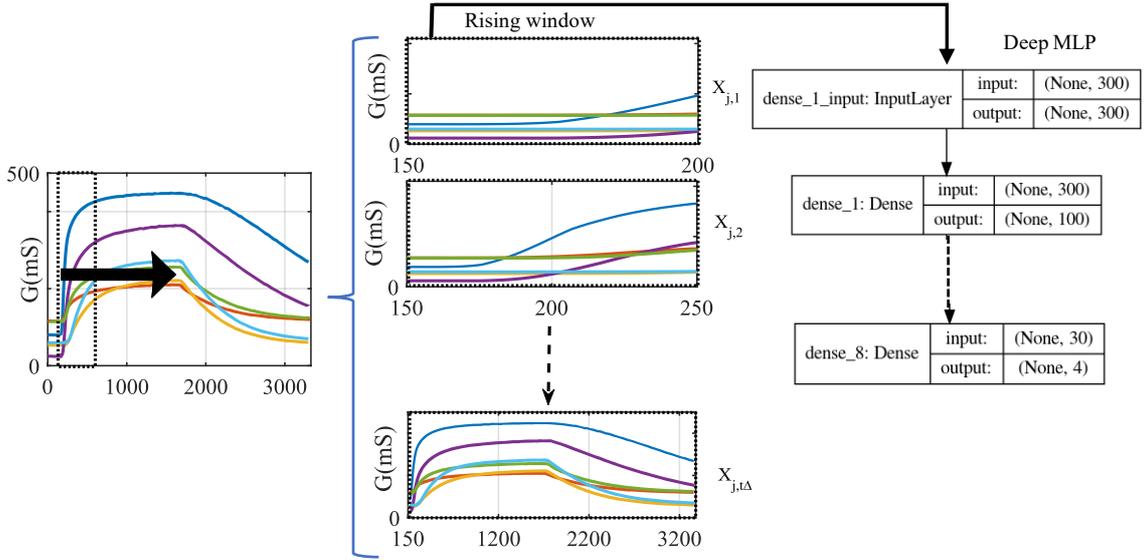

**Fig. 5.** Rapid and online detection approach. Rising window protocol applied to the raw data searching for the minor portion of data to train the deep MLP classifier with the best performance. On the right side is depicted the neural network architecture for the $X_{j,1}$ window with an input size of 300 points and four outputs (three wine spoilage levels and ethanol). The meaning of "None" is unspecified input because we reshaped the data in a flatted array.

The method to find the minor portion of information is as follows. Given the data series $X_j = x_1, x_2, \ldots, x_N$, that represents the gas sensor response $j = 1,2 \ldots, 6$, their corresponding rising windows are defined as: $X_{j,t} = x_{j,1}, \ldots, x_{j,t\Delta}$, where $t = 1, \ldots, \left[\frac{N}{\Delta}\right]$, the step is $\Delta \leq N \wedge \Delta \in \mathbb{N}$, the window size is $t\Delta$, and the operator [.] denotes taking the integer part of the argument. The time series in each window $X_{j,t}$ are used to train the deep MLP classifier. **Fig. 5** exhibits the application of the rising windows protocol in our dataset with $\Delta = 50$, hence each $X_{j,1}$ window has 50 points, each $X_{j,2}$ window has 100 points, and so on. The example architecture of the deep MLP, shown in the same figure, corresponds to the neural network used to process the data for the $X_{j,1}$ window. In this case, the input layer size corresponds to the first window ($t = 1$), six sensors, step $\Delta = 50$; then, it has 6 (1x50) = 300 points. The only data preprocessing applied before the deep MLP neural network was a simple data scaling in each window.

## 3. Results

*3.1 Data exploratory analysis*

We performed the database exploratory analysis using the Principal Components Analysis (PCA). The scores for the first components (2D and 3D plots) for the wines are shown in **Fig. 6.** We also graph the PCA scores of ethanol jointly wines, as shown in **Fig. 7**.

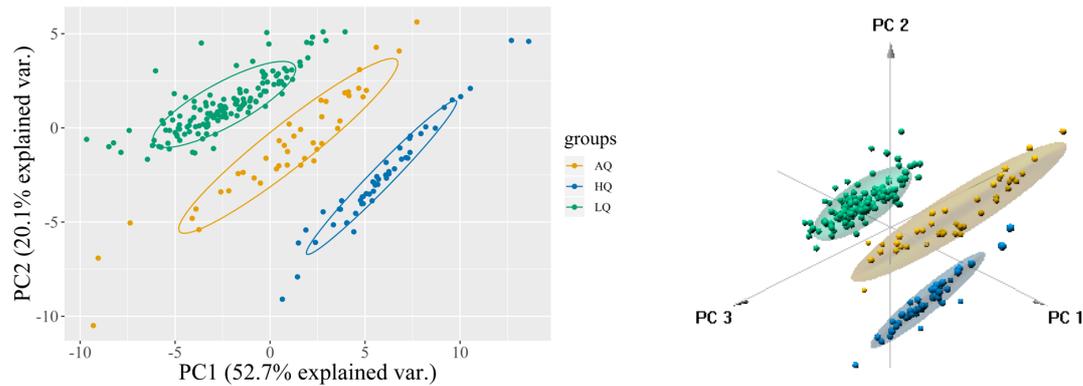

**Fig. 6.** PCA for the three wine groups HQ, AQ, and LQ. On the left side in 2D and the right side in 3D. It is revealed that O-NOSE detects differences between the three groups according to its quality and spoilage threshold. In this case, the three principal components capture a cumulative variance of 81%.

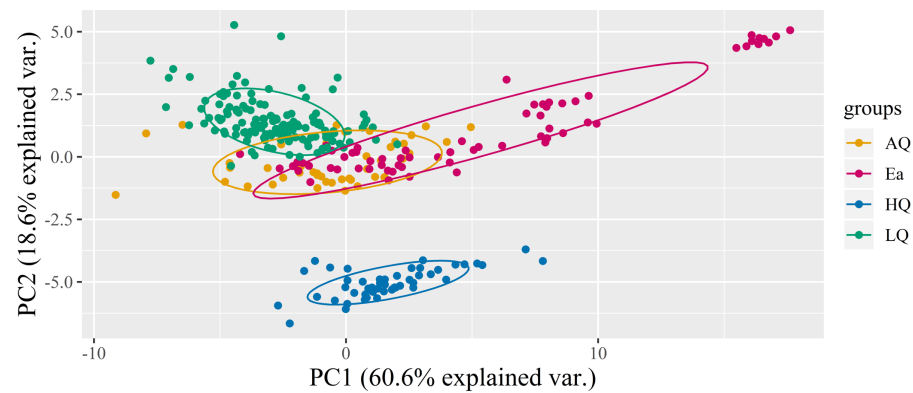

**Fig. 7.** PCA for the three wine groups (HQ, AQ, LQ) and ethanol (Ea). The close relationship between ethanol and wine is evidenced more strongly for the wines labeled as AQ. The groups labeled as HQ and LQ have greater separation regarding the ethanol. In the case of HQ, the organoleptic characteristics are rich in other elements that characterize the excellent taste. For the LQ, the taste is commonly described as vinegar or metallic taste and low level of ethanol.

Based on this exploratory analysis, we performed two experiments with the aim of comparing the performance when the classes are only wines with three spoilage thresholds, and when the ethanol is present as an additional class, which is evidenced as a more complex problem because the ethanol is an essential wine component. These two experiments were performed so much for the conventional approach using SVM, as for the rapid and online detection approach using deep MLP neural network, and the results were compared at the end of Section 3.

*3.2 Conventional approach to classification using SVM*

We used an SVM classifier applying the technique known as Leave One Out (LOO), selecting the measurements of one bottle for the validation group and the remaining for the training group. Since as the dataset contains twenty two bottles, we performed this procedure that quantity of times, and we applied five folds cross-validation technique to prevent the overfitting in the training set. We implemented the scripts for this approach using Matlab R2016a and the Statistics and Machine Learning Toolbox - version 10.2; and, to ensure the integrity of the results, we repeated the procedure 100 times with data shuffling before each training. Then, we averaged the accuracy of each experiment.

In **Table 3** are shown the parameters set on the SVM classifier for the two experiments performed: experiment 1 to discriminate among the three wine thresholds (LQ, AQ, and HQ); and experiment 2 to classify among the three wine thresholds and ethanol (LQ, AQ, HQ, and Ea). The recognition accuracy for training and validation, in the first experiment, was 99.78% and 97.34%, and, for the second experiment, 98.31% and 96.23%, respectively.

**Table 3.** Parameters of the SVM classifiers used for each experiment.

| Parameter | Experiment 1 | Experiment 2 |
| --- | --- | --- |
| Kernel function | Gaussian | Gaussian |
| Kernel parameter scale (gamma) | 8.3 | 19 |
| Box constraint level (C penalty parameter) | 10 | 10 |
| Multiclass method | One-vs-One | One-vs-One |
| Standardize data | True | True |
| Feature selection: variables used in the model | 69 | 56 |
| PCA | disabled | disabled |

*3.3 Rapid and online detection approach using deep MLP*

We did several simulations to find an early portion of the raw data with the best recognition performance in the two experiments. To achieve this, we applied the rising window protocol searching for the minor portion of data to train the deep MLP classifier and averaging the accuracy of each experiment. In this way, we applied the LOO-protocol like the before experiments (Section 3.2), but now training the deep MLP models of eight layers with full neurons connections as detailed in **Table 4** (architecture examples of experiment 2).

**Table 4.** Network architecture of three models for the classification using deep MLP, where $X_{j,t}$ is the time series in each window $t$. The trainable parameters are computed as the multiplication between the inputs and the number of neurons in each layer plus the bias number (see the examples for the layers one and eight in the $X_{j,1}$ model).

| Layer | Neurons | Trainable parameters | | |
|---|---|---|---|---|
| | | $X_{j,1}$ model | $X_{j,12}$ model | $X_{j,63}$ model |
| 1 | 100 | (300x100)+100=30100 | 360100 | 1.8901E+6 |
| 2 | 30 | 3030 | 3030 | 3030 |
| 3 | 30 | 930 | 930 | 930 |
| 4 | 30 | 930 | 930 | 930 |
| 5 | 30 | 930 | 930 | 930 |
| 6 | 30 | 930 | 930 | 930 |
| 7 | 30 | 930 | 930 | 930 |
| 8 | 4 | (30x4)+4=124 | 124 | 124 |

The original raw data have 3330 points, but as was explained in Section 2, the baseline is not considered. Thus, we defined the interval to analyze from the point 150 to the point 3300 (to ensure an integer $\left[\frac{N}{\Delta}\right]$). Since as the step was $\Delta= 50$, we trained 63 models that correspond to each $X_{j,t}$ window using python 3.5.3, repeating the procedure 100 times with data shuffling. In the first experiment with the rapid and online detection approach, the accuracy for the windows with the best performance in the training data was 100%, that occurred 97% of the times in windows with a size less or equal than $X_{j,24}$. This corresponds to the first 64.86 seconds of the raw data interval. In validation data, the accuracy for the windows with the best performance was 97.68%, that occurred 88% of the times in the first window ($X_{j,1}$). This represents only an early portion of the raw data, that is equivalent to the first 2.7s, indicating a significant reduction in the time for the recognition when compared to the conventional approach using the feature extraction/selection method.

The results for the second experiment with this approach indicated that the best performance occurred in windows with a size less or equal than $X_{j,13}$, corresponding to the first 35.13 seconds of the raw data interval. The accuracy was 99.99%, and 96.34%; occurring 54% and 61% of the times in training and validation, respectively. Note that, the separability of the data in this experiment is more complex than the experiment 1 that includes only the three wine spoilage levels, causing that the early portion time necessary for the recognition task being greater. However, it is still less than using the conventional approach which consumes the whole measurement time, suggesting outperformance for the online detection approach using deep MLP.

4. Discussion

The comparison based on the test results between the two discussed approaches is presented in **Table 5**. We highlight the gain in timing for recognition wine quality with our approach, and the possibility of using this approach for online detection without preprocessing techniques.

**Table 5.** Comparison between the conventional and the rapid detection approach.

| Summary of test results | *Conventional approach* | | *Rapid and online detection approach* | |
|---|---|---|---|---|
| | Experiment1 | Experiment2 | Experiment1 | Experiment2 |
| Average accuracy (%) | 97.34±0 | 96.23±0 | 97.68±4.6x10$^{-3}$ | 96.34±4.6x10$^{-3}$ |
| Time for recognition (s) | 171.89 | 171.89 | **2.7** | **35.13** |
| Data preprocessing | FE + FS | FE + FS | Scaling | Scaling |
| Online | NA | NA | Yes | Yes |
| Input size | 69 | 56 | 300 | 3900 |

| Time for training (s) | 16 | 27 | 99 | 130 |
| Time for validation (s) | <<1 | <<1 | <<1 | <<1 |

Average accuracy is presented as the mean ± standard deviation obtained from 100 repetitions. The Mann-Whitney-Wilcoxon test was conducted with (P>0.05). FE: Feature extraction; FS: Feature selection; NA: Not available.

The rapid and online detection approach has the highest computational time in the training. However, the training is performed offline and in most cases is performed just once. Besides, the computational time using the trained model is about a few milliseconds (<<1s) for the two approaches and experiments. Finally, to support the results obtained and assuming independence between both approach with 5% of significance level, we performed the statistical comparison tests. The results revealed that there is enough evidence to say that in the two experiments the accuracy values for the forecasting with the conventional approach is less than the accuracy values for rapid and online detection approach.

In **Table 6**, we compared the results of (Peng, Zhao, Pan, & Ye, 2018) and (Längkvist, Coradeschi, Loutfi, & Balaguru Rayappan, 2013) with our results. We chose these approaches because, unlike the classical feature selection method used in artificial olfactory systems, they also used the raw data to process the gas signals. In that way, in (Peng, Zhao, Pan, & Ye, 2018) was presented an approach based on a Deep Convolutional Neural Network (DCNN) tailored for gas classification but using the entire signal measurement of the gas sensors, resulting in a disadvantage regarding to our approach that lets to achieve faster results using only an early portion of the signals. In (Längkvist, Coradeschi, Loutfi, & Balaguru Rayappan, 2013), similar to the approach proposed in our work, they considered only the transient response centered on an online solution but using unsupervised learning techniques (stacked restricted Boltzmann machines and auto-encoders), although they also focused on obtaining a rapid response, the accuracy of the system is not high. Therefore, our results are better in terms of the time needed to perform the detection. The comparison suggests that it is possible to obtain better results in accuracy and time, using our method. Therefore, our approach is promising for online analyses in E-Nose with low complexity in hardware using standard gas sensors.

**Table 6.** Comparison of the rapid detection approach with other similar works.

|  | (Peng, Zhao, Pan, & Ye, 2018) | (Längkvist, Coradeschi, Lotfi, & Balaguru Rayappan, 2013) | | Proposed work | |
| --- | --- | --- | --- | --- | --- |
|  |  | Result1 | Result2 | Result1 | Result2 |
| Model | DCNN | *DBN* | | Deep MLP | |
| Method | Supervised | *Unsupervised* | | Supervised | |
| Application or gases | CO, $CH_4$, H, and $C_2H_4$ | *Ethanol and TMA* | | Wine samples and ethanol | |
| Gas sensor type | MOS | *Nanostructured ZnO* | | MOS | |
| Online | Not | Yes | | Yes | |
| Average accuracy (%) | 95.2 | 60±4.5 | 83.7±4.1 | 97.68 | 96.34 |
| Time for recognition (s) | 100 | 5 | 25 | 2.7 | 35.13 |
| Time for training (s) | 154 | NA | NA | 99 | 130 |

CO: carbon monoxide; $CH_4$: methane; H: hydrogen; $C_2H_4$: ethylene; TMA: *thrimethylamine;* DBN: *Deep Belief Network;* DCNN: Deep Convolutional Neural Networks; MOS: Metal oxide semiconductor; NA: Not available.

## 5. Conclusions

In this paper, we prove that it is possible to detect wine quality thresholds in a rapid and online way using a deep MLP classifier processing an early portion of the raw data. We obtained an estimation in 2.7 seconds after the gas injection started when we classified three wine spoilage thresholds, and 35.13 seconds when we included ethanol measurements as a class. Therefore, the rapid detection method lets to make predictions 63 times faster for experiment 1, and at least five times faster for experiment 2, when compared with the conventional approach that needs the whole measurement to obtain the main odorant parameters and involves preprocessing techniques.

In this application, we employed Brazilian commercial wines. For future works, it is expected that more researches been conducted including other varieties of wines and more spoilage thresholds. Besides, the rapid detection approach could be extended to other E-Nose applications.

**Acknowledgments**

This work was supported by the Serrapilheira Institute (grant number Serra-1709-22626), Coordenação de Aperfeiçoamento de Pessoal de Nível Superior – Brazil (CAPES) – Finance Code 001, and Conselho Nacional de Desenvolvimento Científico e Tecnológico – CNPq (grant number 420319/2016-6).

**References**

Aleixandre, M., Cabellos, J. M., Arroyo, T., & Horrillo, M. C. (2018). Quantification of Wine Mixtures with an Electronic Nose and a Human Panel. *Frontiers in Bioengineering and Biotechnology*, *6*(February), 1–7. https://doi.org/10.3389/fbioe.2018.00014

Amamcharla, J. K., & Panigrahi, S. (2010). Simultaneous prediction of acetic acidethanol concentrations in their binary mixtures using metalloporphyrin based opto-electronic nose for meat safety applications. *Sensing and Instrumentation for Food Quality and Safety*, *4*(2), 51–60. https://doi.org/10.1007/s11694-010-9092-2

Cretin, B. N., Dubourdieu, D., & Marchal, A. (2018). Influence of ethanol content on sweetness and bitterness perception in dry wines. *LWT - Food Science and Technology*, *87*, 61–66. https://doi.org/10.1016/j.lwt.2017.08.075

De Andrade Lima, L. L., Alexandre, S., Guerra, N. B., Pereira, G. E., De Andrade Lima, T. L., & Rocha, H. (2010). Otimização e validação de método para determinação de ácidos orgânicos em vinhos por cromatografia líquida de alta eficiência. *Quimica Nova*, *33*(5), 1186–1189. https://doi.org/10.1590/S0100-40422010000500032

Gil-Sánchez, L., Soto, J., Martínez-Máñez, R., Garcia-Breijo, E., Ibáñez, J., & Llobet, E. (2011). A novel humid electronic nose combined with an electronic tongue for assessing deterioration of wine. *Sensors and Actuators A: Physical*, *171*(2), 152–158.

Jackson, R. S. (2008). *Wine Science: Principles and Applications*. *Igarss 2014*. Elsevier. https://doi.org/10.1007/s13398-014-0173-7.2

Längkvist, M., Coradeschi, S., Loutfi, A., & Balaguru Rayappan, J. B. (2013). Fast classification of meat spoilage markers using nanostructured ZnO thin films and unsupervised feature learning. *Sensors (Switzerland)*, *13*(2), 1578–1592. https://doi.org/10.3390/s130201578

Lin, X., Yang, F., Zhou, L., Yin, P., Kong, H., Xing, W., … Xu, G. (2012). A support vector machine-

recursive feature elimination feature selection method based on artificial contrast variables and mutual information. *Journal of Chromatography B: Analytical Technologies in the Biomedical and Life Sciences*, *910*, 149–155. https://doi.org/10.1016/j.jchromb.2012.05.020

Lozano, J., Santos, J. P., & Horrillo, M. C. (2016). *Wine Applications With Electronic Noses. Electronic Noses and Tongues in Food Science*. Elsevier Inc. https://doi.org/10.1016/B978-0-12-800243-8.00014-7

Macías, M., Manso, A., Orellana, C., Velasco, H., Caballero, R., & Chamizo, J. (2012). Acetic Acid Detection Threshold in Synthetic Wine Samples of a Portable Electronic Nose. *Sensors*, *13*(12), 208–220. https://doi.org/10.3390/s130100208

Martins, N., Garcia, R., Mendes, D., Costa Freitas, A. M., da Silva, M. G., & Cabrita, M. J. (2018). An ancient winemaking technology: Exploring the volatile composition of amphora wines. *Lwt*, *96*(November 2017), 288–295. https://doi.org/10.1016/j.lwt.2018.05.048

Muezzinoglu, M. K., Vergara, A., Huerta, R., Rulkov, N., Rabinovich, M. I., Selverston, A., & Abarbanel, H. D. I. (2009). Acceleration of chemo-sensory information processing using transient features. *Sensors and Actuators, B: Chemical*, *137*(2), 507–512. https://doi.org/10.1016/j.snb.2008.10.065

Normative instruction N° 14 (2018). Brazil. Retrieved from http://www.agricultura.gov.br/noticias/mapa-atualiza-padroes-de-vinho-uva-e-derivados/INMAPA142018PIQVinhoseDerivados.pdf

OIV. International Organization of vine and wine. Compendium of Internacional Methods of Analysis of wine and Musts, 1 § (2014).

Peng, P., Zhao, X., Pan, X., & Ye, W. (2018). Gas Classification Using Deep Convolutional Neural Networks. *Sensors*, *18*(1), 157. https://doi.org/10.3390/s18010157

Perestrelo, R., Rodriguez, E., & Câmara, J. S. (2017). Impact of storage time and temperature on furanic derivatives formation in wines using microextraction by packed sorbent tandem with ultrahigh pressure liquid chromatography. *LWT - Food Science and Technology*, *76*, 40–47. https://doi.org/10.1016/j.lwt.2016.10.041

Peris, M., & Escuder-Gilabert, L. (2016). Electronic noses and tongues to assess food authenticity and adulteration. *Trends in Food Science and Technology*, *58*, 40–54. https://doi.org/10.1016/j.tifs.2016.10.014

Rodríguez-Méndez, M. L., De Saja, J. A., González-Antón, R., García-Hernández, C., Medina-Plaza, C., Garcíia-Cabezón, C., & Martíin-Pedrosa, F. (2016). Electronic noses and tongues in wine industry.

*Frontiers in Bioengineering and Biotechnology*, *4*(OCT), 81. https://doi.org/10.3389/fbioe.2016.00081

Sáenz-Navajas, M. P., Avizcuri, J. M., Ballester, J., Fernández-Zurbano, P., Ferreira, V., Peyron, D., & Valentin, D. (2015). Sensory-active compounds influencing wine experts' and consumers' perception of red wine intrinsic quality. *LWT - Food Science and Technology*, *60*(1), 400–411. https://doi.org/10.1016/j.lwt.2014.09.026

Stupak, M., Kocourek, V., Kolouchova, I., & Hajslova, J. (2017). Rapid approach for the determination of alcoholic strength and overall quality check of various spirit drinks and wines using GC–MS. *Food Control*, *80*, 307–313. https://doi.org/10.1016/j.foodcont.2017.05.008

Vazallo-Valleumbrocio, G., Medel-Marabolí, M., Peña-Neira, Á., López-Solís, R., & Obreque-Slier, E. (2017). Commercial enological tannins: Characterization and their relative impact on the phenolic and sensory composition of Carménère wine during bottle aging. *LWT - Food Science and Technology*, *83*, 172–183. https://doi.org/10.1016/j.lwt.2017.05.022

Vergara, A., Vembu, S., Ayhan, T., Ryan, M. A., Homer, M. L., & Huerta, R. (2012). Chemical gas sensor drift compensation using classifier ensembles. *Sensors and Actuators B: Chemical*, *166*, 320–329. https://doi.org/10.1016/j.snb.2012.01.074

Yan, J., Guo, X., Duan, S., Jia, P., Wang, L., Peng, C., & Zhang, S. (2015). Electronic Nose Feature Extraction Methods: A Review. *Sensors*, *15*(11), 27804–27831. https://doi.org/10.3390/s151127804

Yan, K., & Zhang, D. (2015). Feature selection and analysis on correlated gas sensor data with recursive feature elimination. *Sensors and Actuators, B: Chemical*, *212*, 353–363. https://doi.org/10.1016/j.snb.2015.02.025

Zhao, Z., Yang, X., Zhao, X., Bai, B., Yao, C., Liu, N., … Zhou, C. (2017). Vortex-assisted dispersive liquid-liquid microextraction for the analysis of major Aspergillus and Penicillium mycotoxins in rice wine by liquid chromatography-tandem mass spectrometry. *Food Control*, *73*, 862–868. https://doi.org/10.1016/j.foodcont.2016.09.035

Zoecklein, B. W., Fugelsang, K. C., Gump, B. H., & Nury, F. S. (1995). Volatile acidity. In *Wine Analysis and Production* (pp. 192–198). Springer, Boston, MA. https://doi.org/10.1007/978-1-4757-6978-4_11